\documentclass[10pt,3p,sort&compress,twocolumn]{elsarticle}

\usepackage{amsmath,amssymb}
\usepackage{epsfig}
\usepackage{graphicx}
\usepackage{bm}
\usepackage{amsmath}

\newcommand{\beq}{\begin{equation}}
\newcommand{\eeq}{\end{equation}}
\newcommand{\beqa}{\begin{eqnarray}}
\newcommand{\eeqa}{\end{eqnarray}}
\newcommand{\beqar}{\begin{eqnarray*}}
\newcommand{\eeqar}{\end{eqnarray*}}

\begin{document}

\begin{frontmatter}

\title{Entropy of extremal black holes from entropy of quasiblack holes}

\author[jpsl]{Jos\'e P. S. Lemos}
\ead{joselemos@ist.utl.pt}
\address[jpsl]{Centro Multidisciplinar de Astrof\'{\i}sica -- CENTRA,
Departamento de F\'{\i}sica,  Instituto Superior T\'ecnico - IST,
Universidade T\'ecnica de Lisboa - UTL,
Av. Rovisco Pais 1, 1049-001 Lisboa, Portugal, \\ \&\\
Institute of Theoretical Physics - ITP, Freie Universit\"at Berlin,
Arnimallee 14 D-14195 Berlin, Germany. \vspace*{.2cm} }
\author[obz]{Oleg B. Zaslavskii}
\ead{ozaslav@kharkov.ua}
\address[obz]{Astronomical Institute of Kharkov V.N. Karazin National
University, 35
Sumskaya St., Kharkov, 61022, Ukraine.}

\begin{abstract}
{The entropy of extremal black holes (BHs) is obtained using a
continuity argument from extremal quasiblack holes (QBHs). It is shown
that there exists a smooth limiting transition in which (i) the system
boundary approaches the extremal Reissner-Nordstr\"{o}m (RN) horizon,
(ii) the temperature at infinity tends to zero and quantum
backreaction remains bounded on the horizon, and (iii) the first law
of thermodynamics is satisfied.  The conclusion is that the entropy
$S$ of extremal QBHs and of extremal BHs can take any non-negative
value, only in particular cases it coincides with $ S=A/4$. The choice
$S=0$ with non-zero temperature at infinity is rejected as physically
unsatisfactory.}
\end{abstract}



\end{frontmatter}

\section{Introduction}

The issue of black hole (BH) entropy is one of the most intriguing in
BH physics. For non-extremal BHs the entropy $S$ is given in terms of
the horizon area $A$ by the Bekenstein-Hawking formula $S=\frac{A
}{4}$ \cite {bekhawk}, a puzzle not yet resolved in fundamental
micro-level terms.  Surprisingly, the issue becomes even more
intriguing in what concerns extremal BHs, as there are two mutually
inconsistent results. There is the prescription $S=0$ obtained from
the fact that for extremal BHs the period of the Euclidean time is not
fixed in a classical calculation of the action \cite{hawteit}, and
there is the usual $S=\frac{A}{4}$ value obtained from string theory
\cite{vafa}. There have been some interesting proposals to further
understand the issue, see \cite{pvisr} for a thermodynamical treatment
and \cite{mpla,lrs} for a semiclassical approach, but the situation
remains contradictory up to now, see \cite{cjr} for the latest
comments. Here, we suggest a resolution of this problem on the basis
of pure thermodynamic arguments. In doing so, we exploit the
quasiblack hole (QBH) approach.

What is the QBH approach? A QBH is a system whose boundary approaches
the would-be horizon as nearly as one likes, and yet the system does
not collapse; a horizon is almost formed but never does
\cite{lwedlezan1lezan2}.  The approach consists in finding the
limiting properties of the system when the boundary tends to its
quasihorizon \cite{lz1lz2lz3lz4lz5}. Properties of such systems are
then compared with pure BH properties. It has been found that, though
worked out through totally different methods, QBHs and pure BHs share
for outside observers the same properties, such as the mass formula
and many others \cite{lz1lz2lz3lz4lz5}, although the interior of both
systems is totally different, interior made of matter for QBHs, vacuum
interior for pure BHs.

In the work \cite{lzent} important developments on QBH properties were
advanced. The entropy of non-extremal QBH systems was found
thermodynamically and shown to be equal to the BH entropy
$S=\frac{A}{4}$.  This was achieved by using on one hand QBH
procedures, and on the other hand the formalism for gravitating
systems, such as BHs, of Brown and York \cite {byby90} for the
definition of quasilocal energy and other quasilocal thermodynamic
quantities. Now, QBHs can be obtained from a quite generic class of
systems, but a simple realization of them is provided through thin
charged shells when these are brought to their own horizon radius. In
the works \cite{martdfpage} these systems were thermodynamically
studied but the analysis fell short of letting the shell approach the
horizon. In \cite {pvisr} a specially arranged shell system was
imagined in order to analyze its behavior when lowering
quasistatically the shell into its own horizon, with the shell being
immersed in an appropriate quantum vacuum. It was found that
$S=\frac{A}{4}$, as well. Our QBH approach for non-extremal systems
\cite{lzent} is generic and thus has none of the drawbacks of
specialization to simple thin shell systems.

Since many properties, in particular the entropy $S$ of non-extremal
QBHs, can be found and match the corresponding quantities of pure BHs,
we continue our pursue and want to shed light on the entropy of
extremal BHs by studying the entropy of QBHs, more specifically,
extremal QBHs. One can then consider a sequence of stars made of some
sort of usual matter, each member of the sequence with lesser radius
say, in which the last member of the sequence is an extremal QBH. By
using stars made of matter one is enabled to consider more prosaic
systems, i.e., systems that do not possess a horizon, and study them
through the usual textbook formalism of thermodynamics. Only for the
last members of the sequence of stars one takes the limit to the
transition to the QBH state. At this very last stage of the sequence,
in order to have a well defined thermodynamic system, one has to use
results from quantum field theory in curved background
\cite{and1and2fn}. The sequential procedure is of immense importance,
as due to the continuity of the calculation process, the QBH approach
enables to evade difficulties connected with the often invoked
potential discontinuity between non-extremal and extremal BHs. Using
the QBH, we follow the procedure developed in \cite{lzent} (see also
\cite{pvisr}), i.e., we calculate $S$ of the material system when the
would-be horizon is approached. Since for an external observer a QBH
and a BH cannot be distinguished \cite {lz1lz2lz3lz4lz5}, one expects
that the entropy of such a system without a true horizon tends to the
entropy of the corresponding BH. In this sense our calculations not
only give an answer for the QBH entropy but elucidate the value of the
entropy of a BH to which the exterior of a QBH tends due to the
continuity of the process in the latter case.

We show that our consistent thermodynamic treatment rejects definitely
the choice $S=0$ but does not give an unambiguous universal result for
$S$. The entropy depends on the properties of the working material
and, moreover, on the manner the temperature approaches the zero
value. In particular $S=\frac{ A}{4}$ is not singled out beforehand
for the extremal BH entropy.

\section{Basic formulas}

The study of extremal QBHs has one advantage over the study of
non-extremal ones. While non-extremal QBHs show a sort of singular
behavior at the quasihorizon, such as a singular stress-energy tensor,
extremal QBHs are nonsingular well behaved systems throughout
\cite{lz1lz2lz3lz4lz5}. In order to make the problem tractable we
stick to spherically symmetric systems.

Consider a spherical symmetric compact body with boundary at $r=R$
such that $r< R$ defines the inner region, $r> R$ the outer one, and
its total charge $ q$ is equal to its ADM mass $m$. The generic
space-time line element in the usual coordinates $(t,r,\theta,\phi)$
is then
\begin{equation}
ds^{2}=-V\exp\,(2\psi) \,dt^{2}+\frac{dr^{2}}{V}+r^{2}\left( d\theta
^{2}+\sin ^{2}\theta \,d\phi ^{2}\right) \,,\label{out}
\end{equation}
where $V$ and $\psi$ are functions of $r$. In general one also needs
an expression for the electric potential $\phi(r)$. For $r\geq R$ the
space-time is described by the extremal Reissner-Nordstr\"{o}m (RN)
metric, in which case $\psi(r)=0$, and by the Coulomb electric
potential,
\begin{equation}
V(r)=\left( 1-\frac{r_{+}}{r}\right) ^{2}\,, \quad \phi
(r)=\frac{r_{+}}{r}+ \mathrm{constant}\,, \label{uv}
\end{equation}
where $r_{+}=m=q$, $r_{+}$ being the gravitational radius of the body,
i.e., the radius of the would-be horizon, and the constant can be
chosen in convenient terms. $R\geq r_{+}$ always holds here and at
$R=r_{+}$ a QBH forms.

The whole system, compact body plus spacetime, is assumed to be in
thermodynamic equilibrium at some non-zero temperature. The entropy of
the system is calculated from integrating the first law when it
undergoes a reversible process. In general, the integration requires
knowledge of the matter equation of state, but we show that, when one
deals with systems on the threshold of forming an extremal horizon,
deep conclusions can be drawn without that knowledge. The first law of
thermodynamics for our system can be written as \cite{byby90}
\begin{equation}
TdS=dE+\lambda\, dA-\varphi\, de\,.  \label{1}
\end{equation}
We now go through $TdS$, $dE$, $\lambda\, dA$, and $-\,\varphi\,de$
carefully.

The local temperature $T$ on the boundary $R$ is related to the
temperature $ T_{0}$ at infinity by the Tolman formula
\begin{equation}
T=\frac{T_{0}}{\sqrt{V(R)}}=\frac{T_{0}}{1-\frac{r_{+}}{R}}\,.
\label{t}
\end{equation}
Since $S$ is the entropy of the system, $dS$ is the change of the entropy
upon changing the other quantities.

The quasilocal energy $E$ is given by \cite{byby90}, $E=R\left(
1-\sqrt{V(R)} \right) $. It is seen from this and (\ref{out}) that for
our extremal system $E=r_{+} $ does not depend on $R$. Thus
\begin{equation}
dE=dr_{+}\,.  \label{dE}
\end{equation}

The gravitational pressure $\lambda $ is found from the inner
region. For the inner region $r\leq R$ the metric is given as in
Eq.~(\ref{out}). Then the gravitational pressure $\lambda $ at the
boundary at $r=R$ equals to \cite{byby90} $8\pi \lambda
=\frac{1}{R}\left( \sqrt{V(R)}-1\right) +\left(
\frac{1}{2}\frac{1}{\sqrt{V(r)}}\frac{dV(r)}{dr}+\sqrt{V(r)}\;\frac{d\psi
(r) }{dr}\right) _{r=R_{-}}\,\text{,}$ where $r=R_{-}$ means that the
derivatives should be taken from the inner region. It follows from the
$tt$ and $rr$ Einstein equations for the inner region and the boundary
condition $ \psi (R)=0$ (mandatory for a smooth matching with the
outer region) that $ \psi =4\pi
\int_{R}^{r}d\bar{r}\,\frac{\bar{r}(p_{r}(\bar{r})+\rho (\bar{r}))
}{V(\bar{r})}$. Here $p_{r}$ is the radial pressure and $\rho $ is the
energy density, both include contributions from the matter and the
electromagnetic field, i.e.,
$p_{r}=p_{r}^{\mathrm{\,matter}}+p_{r}^{\mathrm{ \ \,em}}$ and $\rho
=\rho ^{\mathrm{\,matter}}+\rho ^{\mathrm{\,em}}$. It also follows
from the $tt$ Einstein equation that $V(r)=1-\frac{2m(r)}{r}$, with
$m(r)=4\pi \int_{0}^{r}d\bar{r}\,\bar{r}^{2}\rho (\bar{r})$. Then, for
our concrete system, using (\ref{uv}) and the equation for $\lambda $,
one finds after some manipulations, $8\pi \lambda R=4\pi \,{\
p_{r}^{\mathrm{ \,matter}}R^{2}}/(1-{r_{+}}/{R})$. Now, to make
progress we have to understand the system at the threshold of being a
QBH. We have to take into account that on the quasihorizon
$p_{r}^{\mathrm{\,matter}}(r_{+})=0$ according to our general results
on pressure in \cite{lz1lz2lz3lz4lz5}. When matter is absent in the
inner region, as in a thin shell, this condition is exact. When there
is matter, one can write quite generally $p_{r}^{\mathrm{\
\,matter}}(r)=\frac{b(r_{+},R)}{4\pi R^{2}}\left(
1-\frac{r_{+}}{R}\right) $, valid near $R=r_{+}$ and with the function
$b(r_{+},R)$ model-dependent.  Note that we do not need to impose that
$p_{r}(R)=0$ for $R>r_{+}$, the surface can move due to thermal motion
or something else, or even be a cold star in which case
$p_{r}(R)=0$. The point is that if the body is sufficiently compressed
it follows that $p_{r}^{\mathrm{\,matter}}(r_{+})=0$
\cite{lz1lz2lz3lz4lz5}. Thus, finally,
\begin{equation}
\lambda =\frac{1}{8\,\pi }\,\frac{b(r_{+},R)}{R}\,. 
\label{lambdafin}
\end{equation}
On the other hand, the area $A$ is defined as $A=4\pi R^{2}$, so that
\begin{equation}
dA=8\pi R\,dR\,.  \label{areain}
\end{equation}

The electric potential $\varphi $ represents the difference in
electrostatic potential between a reference point with potential $\phi
_{0}$ and the boundary $R$ with potential $\phi (R)=q/R$, blue-shifted
from infinity to $R$ through the factor $1/\sqrt{V}$, where $V$ is the
time component of the static metric. Thus,
\begin{equation}
\varphi =\frac{\phi _{0}-\phi (R)}{\sqrt{V(R)}}\,.  \label{pot}
\end{equation}
For $de$, since at the quasihorizon limit $e=r_{+}$, one has
\begin{equation}
de=dr_{+}  \label{de}
\end{equation}
We are now ready to analyze the entropy of quasiblack holes.

\section{Entropy of quasiblack holes and entropy of extremal black holes}

First, let us consider the simplest case: a charged shell with a flat
space-time inside and an extremal RN metric outside. Then, $b=0$ since
there is no matter inside. Also, as the potential is everywhere
constant inside, one has $\varphi =0$. Then, we obtain the first law
in the form,
\begin{equation}
TdS=dr_{+}\text{.}  \label{1s}
\end{equation}
It is instructive to recall that in the case of uncharged shells, as
treated in \cite{martdfpage}, the integrability condition of the first
law yields $ T_{0}=T_{0}(r_{+})$, so $T_{0}$ is not a function of $R$
in such a case. Now our case is an extremal charged shell rather than
an uncharged one. In this case the integrability condition for
Eq.~(\ref{1s}) is $T=T(r_{+})$, i.e., the local temperature is a
function of $r_{+}$ alone. On the other hand the temperature at
infinity has thus the form,
\begin{equation}
T_{0}=T(r_{+})\left(1-\frac{r_{+}}{R}\right)\,,  \label{t0s}
\end{equation}
It contains a dependence on $R$, but, as usual, it does not depend on
$r$.  With these remarks we can now integrate Eq.~(\ref{1s}) and
obtain
\begin{equation}
S=S(r_{+})=\int_{0}^{r_{+}}d\bar{r}_{+}\,\frac{1}{T(\bar{r}_{+})} \,,
\label{entshell}
\end{equation}
where the constant of integration ensures that $S\rightarrow 0$ when the
system shrinks to nothing. To be sure, Eq.~(\ref{entshell}) is valid for any
$R\geq r_{+}$.

Second, we consider a more general configuration, with the inside
having some type or another of distribution of matter other than
vacuum. Clearly, one has to assume that the integrability conditions
for the system are valid, otherwise there is no thermodynamic
system. Then, since $S$ is a total differential one can integrate
along any path. Choose the path $R = r_{+} (1 + \delta)$ with $\delta$
constant and small, so that $dS=\mathrm{ (something)}\,dr_{+}$. Then
one can integrate this equation to obtain $S$.  Taking then at once
the limit $R\rightarrow r_{+}$, we obtain instead of (\ref{entshell})
the following equation,
\begin{equation}
S=S(r_{+})=\int_{0}^{r_{+}} d\bar{r}_{+}\,\frac{D(\bar r_{+})}{T(\bar{r}
_{+}) } \,,  \label{entropy}
\end{equation}
where,
\begin{equation}
D(r_{+})= 1+b_{+}-\varphi _{+}\,,  \label{entropy2}
\end{equation}
$b_{+}=b(r_{+}, R=r_{+})$ and $\varphi _{+}=\varphi
_{+}(r_{+},R=r_{+}) $.  In general, we only require $1+b_{+}-\varphi
_{+}>0$ to ensure the positivity of the entropy. Note that if the
density of matter inside vanishes at $r=R$, we return to the thin
shell situation, since $b_{+}\rightarrow 0$, $\varphi _{+}\rightarrow
0$, and so $D(r_{+})=1$.

Thus, we can state the following. For QBHs, for any finite generic
$T(r_{+})$, one obtains a well-defined positive entropy, $S>0$, from
(\ref{entropy}), as well as a vanishing temperature at infinity,
$T_{0}\to0$, from (\ref{t0s}).  In addition one can consider the case
in which $T(r_{+})$ is not finite, $T(r_{+})\to\infty$ as
$T({r_{+}})=\left(T_0/\left(1-\frac{r_{+}}{R} \right)\right)|_{R\to
r_+}$. In this particular instance one obtains from (\ref{entropy})
that for QBHs $S=0$ and from (\ref{t0s}) that $T_{0}$ is positive and
finite, not equal to zero, $T_{0}>0$. This latter particular case of
QBH behavior is equivalent to the prescription given in \cite
{hawteit} for extremal pure BHs.

We now argue that for extremal pure BHs the prescription $T_{0}\neq 0$
of \cite{hawteit} is unsatisfactory. As pointed in \cite{and1and2fn},
the prescription that $T_{0}\neq 0$ is an arbitrary finite quantity,
is inconsistent with quantum backreaction. Indeed, the corresponding
quantum stress-energy tensor is of the form
${T^{\mathrm{quant}}\,}_{\mu }^{\nu }=T^{4}f_{\mu }^{\nu }+h_{\mu
}^{\nu }$ where $h_{\mu }^{\nu }$ is a term finite everywhere. Near
the horizon the first term of ${T^{\mathrm{quant}}\,} _{\mu }^{\nu }$
diverges as the local temperature $T$ diverges due to the redshift
factor. This unstabilizes the system and is physically inappropriate
at the semi-classical level \cite{and1and2fn}. Actually, if this were
true, the temperature of the quantum fields and that of the BH itself
would not coincide, making thermal equilibrium impossible. $T_{0}\neq
0$ and $S=0$ cannot be a solution. One is left with vanishing $T_{0}$
($T$ finite) and $S>0$ undetermined for extremal BHs.

Our QBH approach gives consistency to this solution of the
thermodynamic extremal BH problem. Indeed, the result provided by
Eqs.~(\ref{t0s}) and~(\ref{entropy}) is free of difficulties. As the
local temperature $T(r_{+})$ remains finite when $R\rightarrow r_{+}$,
the quantum stress-energy tensor ${T^{\mathrm{quant}}\,}_{\mu }^{\nu
}$ on the quasihorizon remains finite or even negligible. Moreover,
the first law of thermodynamics is also satisfied with the choice
(\ref{entropy}). Thus, thermal equilibrium is kept in the system, the
temperature tends to the Hawking value with a suitable rate, given by
(\ref{t0s}), and $S> 0$ is somehow undetermined. Can we
nevertheless say something more definite about the form of the
function $ S(r_{+})$? Eq.~(\ref{entropy}) tell us that the situation
is model-dependent, it depends on $D(r_{+})/T(r_{+})$, which depends
on the properties of the particular system under study. For instance,
only for special cases, when the quantity $D(r_{+})/T(r_{+})$ is given
by $ D(r_{+})/T(r_{+})=2\pi r_{+}$, can we obtain the
Bekenstein-Hawking value $ \frac{A}{4}$ where $A$ is the area of the
quasihorizon surface. In addition, for a given model, changing the
parameter $T(r_{+})$, say, one can obtain any desirable value for $S$,
with $S>0$, $S=0$ being ruled out.

In deriving that the entropy of extremal BHs is model-dependent we are
not alone. We were preceded by the results of \cite{pvisr}. In
\cite{pvisr} particular thin shells as working material were analyzed,
and the Gibbs-Duhem relation (which for self-gravitating systems is,
in general, not valid) was used, to support the conclusion that
extremal BH entropy is model-dependent. Our approach is much more
general, makes no use of thin shells neither of the Gibbs-Duhem
relation. Moreover, in deriving that the manner in which the
temperature $T_{0}$ approaches zero is not well fixed, as $T(r_{+})$
is a free quantity, we are also not alone. We were preceded by the
results of \cite{mpla} and \cite{lrs}. Indeed, remarkably, on a
totally different setting and actually in a work which raised for the
first time problems connected to extremal BHs alone, it was shown in
\cite{mpla} that at the extremal state fluctuations on the temperature
grow unbound. Our work shows the appearance of unusual features in the
thermal description even without considering such fluctuations. This
problem is a quite separate non-trivial issue needing further
consideration. In addition, \cite{lrs} has concluded that the notion
of zero temperature is ill-defined for extremal BHs, whereas we
defined it but in a rather delicate way (see Eq.~(\ref{t0s})), so it
changes when we go through the referred sequence of configurations.

\section{Conclusions}
We have obtained the expression for the entropy in Eq.~(\ref{entropy})
(see also Eq.~(\ref{entshell})).  This expression is valid for any
$R>r_{+}$. We have been interested in the quasiblack hole limit
$R\rightarrow r_{+}$ in the course of which the temperature at
infinity $T_{0}$ obeys $T_{0}\rightarrow 0$ according to
Eq.~(\ref{t0s}). In this regard, we want to emphasize the difference
between the system under discussion and traditional thermodynamics. In
the latter, the state is characterized by its thermodynamic parameters
with no memory on how their values were achieved. Therefore, in the
limit when $T_{0}\rightarrow 0$ and $R\rightarrow r_{+}$ one could
naively expect to obtain some unambiguous quantity for $S$
corresponding to $R=r_{+}$, $T_{0}=0$. Instead, our approach implies
either that the entropy of extremal QBHS, and by inference of extremal
BHs, is not a full-fledged unambiguous quantity, in the sense that any
desirable value of $S$ can be achieved by tuning $T(r_{+})$ say, or
that $T(r_{+})$ is unique and can be found on fundamental grounds in a
semiclassical theory. One should verify this hypothesis. In any case,
our work shows that near $T_0=0$, i.e., near the extremal QBH or
extremal BH limit, the usual thermodynamic picture can change
drastically.  In particular, the fact that we cannot simply take the
limit $ T_{0}\rightarrow 0$ but, instead, should consider different
ways of its approaching to zero depending on $T(r_{+})$ as in
Eq.~(\ref{t0s}), makes this issue much more intricate than expected.

We stressed the key role played by the QBH concept and have shown how
to substantiate the choice for the extremal BH entropy from a
thermodynamic stand. The result is not universal, with $S=0$,
$T_{0}\neq 0$ being ruled out. We used continuity arguments and so one
question we should ask is whether the limiting configuration in the
QBH setup yields an entropy $S$ that can be consider the entropy of an
extremal BH. Our approach stems from taking the horizon limit of
matter configurations with time-like boundaries, whereas BHs have from
the start a lightlike horizon. Can we trust that by continuity from
the QBH approach we get the correct entropy of an extremal BH?
Non-extremal QBHs yield to continuity arguments \cite{lzent}, but
there one knew the result beforehand. However, now the entropy of an
extremal BH is unknown, so there is no gauge to compare with. Thus,
the situation is more tricky, and though we do not possess a rigorous
proof, we can add arguments in its favor. When we change $m$ and $q$,
approaching the extremal RN BH metric from a non-extremal one, jumps
in $S$ are not excluded.  However, these jumps should be connected
with jumps in the temperature. If we take the prescription of
\cite{hawteit}, $T_{0}$ changes from $ T_{0}\approx 0$ for the
near-extremal configuration to finite $T_{0}$. In contrast, in our QBH
approach $T_{0}\rightarrow 0$ smoothly with no source of
discontinuity. Moreover, using the standard approach for the entropy
of an extremal BH, there remains the difficulty of its calculation and
definition within thermodynamics. If one takes the prescription of
\cite {hawteit}, $T_{0}$ is finite and arbitrary, but backreaction
destroys the horizon. If, instead, one puts $T_{0}$ to zero in
accordance with its Hawking value, it is not quite clear how to obtain
an entropy by differentiating the system's free energy with respect to
a fixed zero temperature. On the other hand, the QBH approach evades
these problems since a horizon is absent and at each stage it has a
well-defined small non-zero $ T_{0}$. It seems appropriate to consider
the limiting entropy of the sequence of the QBH configurations
precisely as a definition of extremal BH entropy, analogously to the
operational definition substantiated in \cite {pvisr}.

\section*{Acknowledgments}

This work was funded by FCT - Portugal, through projects
CERN/FP/109276/2009 and PTDC/FIS/098962/2008. JPSL also thanks the FCT
grant SFRH/BSAB/987/2010.  The work of O.Z. was supported in part by
the ``Cosmomicrophysics'' programme of the Physics and Astronomy
Division of the National Academy of Sciences of Ukraine.

\end{document}